# Hybrid plasmonic Bound State in the Continuum entering the zeptomolar biodetection range

Elena Clabassi[∥,1,2], Gianluca Balestra[∥,1,2], Giulia Siciliano[1], Laura Polimeno[1], Iolena Tarantini[2], Elisabetta Primiceri[1], David Maria Tobaldi[1], Massimo Cuscunà[1], Fabio Quaranta[3], Adriana Passaseo[1], Alberto Rainer[1,5], Silvia Romano[4], Gianluigi Zito[4], Giuseppe Gigli[1,2], Vittorianna Tasco[1,‡] and Marco Esposito[1,2]*

1. CNR NANOTEC Institute of Nanotechnology, Via Monteroni, Lecce 73100, Italy

2. Department of Mathematics and Physics Ennio De Giorgi, University of Salento, Lecce 73100, Italy

3. CNR IMM Institute for Microelectronics and Microsystems, Via Monteroni, Lecce 73100, Italy

4. CNR ISASI Institute of Applied Sciences and Intelligent Systems, Naples 80078, Italy

5. Department of Engineering, University Campus Bio-Medico di Roma, via Álvaro del Portillo 21, 00128 Rome, Italy

*E-mail: marco.esposito@nanotec.cnr.it

∥E.C. and G.B. contributed equally to this work.

‡ Vittorianna Tasco is currently seconded at the European Research Council Executive Agency of the European Commission. Her views expressed in this paper are purely those of the writer, may not in any circumstance be regarded as stating an official position of the European Commission.

ABSTRACT

Optical Bound States in the Continuum are peculiar localized states within the continuous spectrum that are unaffected by any far-field radiation and intrinsic absorption, therefore possessing infinite mode lifetime and Q-factor. To date they have been widely studied in dielectric structures whereas their exploitation in lossy media, i.e. plasmonic nanostructures, still remains a challenge. Here, we show the emergence of a hybrid BIC state in a 2D system of silver-filled dimers, quasi-embedded in a high-index dielectric waveguide. The hybrid BIC onset is found to be highly dependent on the bare modes' spectral and spatial overlap, but particularly on the plasmonic field's intensity. By tailoring the hybridizing plasmonic/photonic fractions we select an ideal coupling regime for which the mode exhibits both, high Q-factor values and strong near-field enhancement tightly confined in the nanogap and a consequently extremely small modal volume. We demonstrate that this optical layout can be exploited in a proof-of-concept experiment for the detection of TAR DNA-binding protein 43, which outperforms the sensitivity of current label-free biosensing platforms, reaching the zeptomolar range of concentration.

KEYWORDS

hybrid metasurfaces, bound states in the continuum, localized surface plasmon resonances, hybrid BICs, zeptomolar biosensing

INTRODUCTION

In photonics, Bound States in the Continuum (BICs) have garnered a great deal of interest due to their appealing features. BICs have been proved to be a novel method of light confinement, coexisting within the continuous spectrum of radiating waves, but remaining inaccessible to the far-field. This is why BICs are additionally referred to as embedded trapped modes characterized by potentially infinite quality factor and particularly narrow spectral linewidth [1,2]. Symmetry-protected (SP) BICs have received by far the greatest attention out of all the different kinds of BICs: they emerge at extremely symmetric points, namely the Γ-point in the reciprocal space of a periodic structure, and cannot couple to leaking modes because of a symmetry mismatch. The structural symmetry also makes these states robust against defects or perturbations.

Even though BICs are modes in the radiation continuum, the lack of radiative losses have made them a promising tool in a variety of applications such as lasing [3,4], filtering [5] and sensing. Notably, in this latter field of interest, BIC-based optical solutions that use innovative structural architectures towards high sensitivity are emerging [6,7], even though they have not yet demonstrated detection below the femtomolar concentration levels of target molecules [8–13].

To enhance the performance in light-matter interactions of potential application in biosensing, the hybridization of different modes can be considered [14,15]. For example, pure plasmonic modes can confine light at a subwavelength scale but their resonances are characterized by low Q-factor values, due to significant radiation and ohmic losses. Hybridization of plasmonic and photonic modes demonstrated to overcome this drawback [14–16]. Similarly, hybrid photonic–plasmonic BICs (hybrid BICs) can be considered, as in the recently reported 1D and 2D systems [17–24].

Advancing the state-of-the-art, we show the emergence of an hybrid SP-BIC in the visible range within a 2D metasurface platform consisting of a silver-filled periodic dimers quasi-embedded in a silicon nitride ($Si_3N_4$) waveguide, where the hybrid BIC trigger is polarization-driven and is affected by the detuning energy (DE) between the guided modes and the LSP resonance.

Our nanosystem enables multifunctional properties exploiting intrinsic configurational features that empower high Q/V values and a strong electromagnetic (EM) field enhancement when incident light is polarized along the dimers, both amplified by the nanogap effect. The proposed platform is engineered to be set in a region of DE that corresponds to the optimal trade-off between high Q values and small modal volumes by finely tuning of the plasmonic/photonic fractions that govern the coupling mechanism. by LP scanning.

Moreover, the 2D features of the system grant direct access to the active surface for light-matter applications, in particular biosensing. Indeed, the system, in combination with a bio-inspired reactive coating, was exploited for the detection of a biomarker associated with neurodegenerative diseases, the transactive response (TAR) DNA-binding protein 43 (TDP-43), allowing its optical detection up to zeptomolar concentrations. To the best of our knowledge, such a high sensitivity represents the first biosensing signature reaching this detection level exploiting BIC technology.

RESULTS

To demonstrate the optical emergence of a plasmonic symmetry-protected BIC we numerically and experimentally developed a new paradigm of a quasi-embedded metasurface consisting of a square array of fully silver-filled nanoholes dimers inside a $Si_3N_4$ waveguide slab (Fig. 1) (see experimental section). This choice can be particularly effective in view of biosensing applications since the dimers design allows to achieve enhanced plasmonic field (within the nanogap), while the surface air exposure enables easy access to the plasmonic field and maximizes its superposition with biomaterials. In this matter, it is noteworthy that present state-of-the-art designs are limited to plasmonic components either covered entirely by dielectric claddings[18,19] or placed on top of a waveguide layer[21,22].

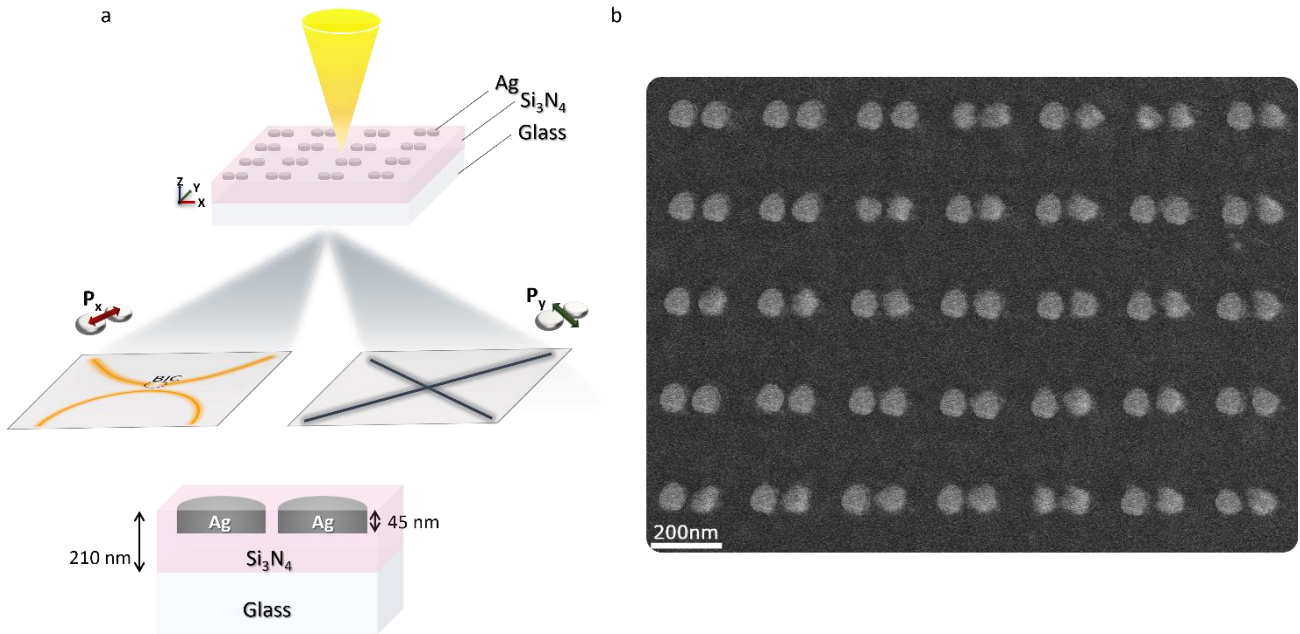

*Figure 1. a) Schematic view of the dimer based nanoarchitecture supporting hybrid BIC generation in a switchable configuration depending on incident light polarization. b) SEM image (top view) of the fabricated dimer array embedded in the $Si_3N_4$ waveguide. The square arrays of silver ND dimers (70 nm ND diameter, 320 nm lattice period along x-axis, 300 nm lattice period along y-axis, 45 nm thickness and 30 nm gap) have been realized inside a $Si_3N_4$ slab of 210 nm thickness by a combination of PECVD, electron beam lithography, ICP RIE and thermal evaporation, as detailed in the experimental section.*

Design of dimers-based nanoarchitecture for hybrid SP-BIC generation

A non-radiative BIC shows up at the gamma point of a 2D dielectric array when it has subdiffractive lattices and the nanoparticles support only the in-phase resonant out-of-plane dipole mode[25]. These oscillating dipoles along the normal incident direction do not radiate towards it and no radiation from the system is allowed due to the destructive interference in all other directions [26].

It is possible to create a radiation channel that transforms the BIC mode from a purely subdiffractive state into a leaky resonance (quasi-BIC) with a high finite Q-factor. This can be achieved by precisely adjusting one of the periods to support a diffraction order at the wavelength of the nanoobject's resonance, or by establishing a refractive index contrast around the nanoparticles, such as embedding them within a high-index dielectric waveguide. [23].

In our case, following the latter approach (dielectric waveguide in silicon nitride), when incident light hits the sub-periodically modulated structure, it couples to guided modes, causing the appearance of the resonant features in the spectra and following the dispersion curves of the guided modes in the Fourier space. Via Fourier-plane spectroscopy, we numerically investigated the effects of a 2D holes-based periodic grating on the fundamental guided modes in the slab and found that the coupling between these and the first diffractive orders leads to the bands' anticrossing and BIC emergence for both incident polarizations (Fig. S1).

Moving away from the BIC position, the non-zero radiative losses transform the BIC in a quasi-BIC with high Q-factor. Up to now, these optical quasi states have been used for various applications. This is because the high Q value balances the large mode volume V (low field enhancement M) of the all-dielectric nanostructure resonance in the Q/V ratio. This ratio defines the metrics for nanoscale enhancement[27]. Anyway, to reach high Purcell factor values, the electromagnetic field has to be strongly enhanced. This motivates the research for hybrid guided plasmonic modes where the optical near field can be strongly enhanced by a subwavelength plasmonic component while still preserving low-losses high Q factor characteristics of the dielectric counterpart [21]. For this purpose, we filled the dimers-arranged nanoholes with silver. The metal's choice is driven by the possibility to observe its optical response in the visible range (Fig. S2). The LSP resonances sustained by the metallic nanostructures enable the coupling with the BIC-like photonic branches leading to modes splitting and formation of a hybrid plasmonic BIC mode. Moreover, this design offers flexibility through structural parameters (i.e. NDs diameter, lattice periods, nanogap's dimensions, slab thickness) and incident light polarization (with respect to the dimer axis).

In order to gain insight into the features of the hybrid BICs, we plot in Fig. 2a-d the measured and simulated angular dispersion reflection maps as a function of the incident angle ($\theta°$) for incident light polarization parallel ($P_x$) and normal ($P_y$) to the dimer axis.

Along the $P_x$ polarization (Fig. 2a), we noted the emergence of the symmetry-protected hybrid BIC at 2.2 eV triggered by the dimers LSP signature that is redshifted with respect to the photonic modes (Fig. S2a). In this case, although the DE between the modes is non-zero, the spatial overlap and intensity of the electric fields involved, namely the plasmonic field in the nanogap and the photonic BIC's one, lead to hybrid BIC generation. The physical origin of the BIC state is further identified by the corresponding electric and magnetic field distributions calculated at the quasi-BIC point and displayed in Fig. 2c,d as it has been already demonstrated in the literature: the electric field vectors on the xy plane (perpendicular to the excitation plane) come out of the dimers while the magnetic field vectors (extracted from the z-plane) rotate around the structures[20]. Switching to the $P_y$ polarization (Fig. 2b), the LSP feature is centred at 2.3 eV (Fig. S2b) and refers to the single ND, spectrally overlapping with the photonic modes. Here, the map shows dark crossing modes, indicating that no BIC onset is triggered, due to reduced electric field strength of the single ND. The near electric field distributions calculated at the two considered polarizations are displayed in the insets of Fig. 2e,f and allow to highlight the effect of the dimer gap as we move from the $P_x$ case, where the field is more confined inside the gap, to the Py polarization where it is highly delocalized and around the dipolar lobes of the two disks. Also, the intensity value of the field is remarkably enhanced along $P_x$ with respect to the opposite polarization, sustaining the claim that in the latter is too weak to trigger the hybrid BIC generation. Following this analysis, we can argue that in strong coupling regime characterized by limited detuning energy the high LSP losses act as an additional destructive leaky channel overcoming the Rabi Splitting $\Omega_R < \frac{(\gamma_{LSP}-\gamma_{GM})}{2}$, where $\gamma_{LSP}$ and $\gamma_{GM}$ represent the LSP and guided modes losses, respectively, . This issue is compounded by the fact that in this case the electric fields (plasmonic and photonic) may not be spatially overlapping and thus they do not couple.

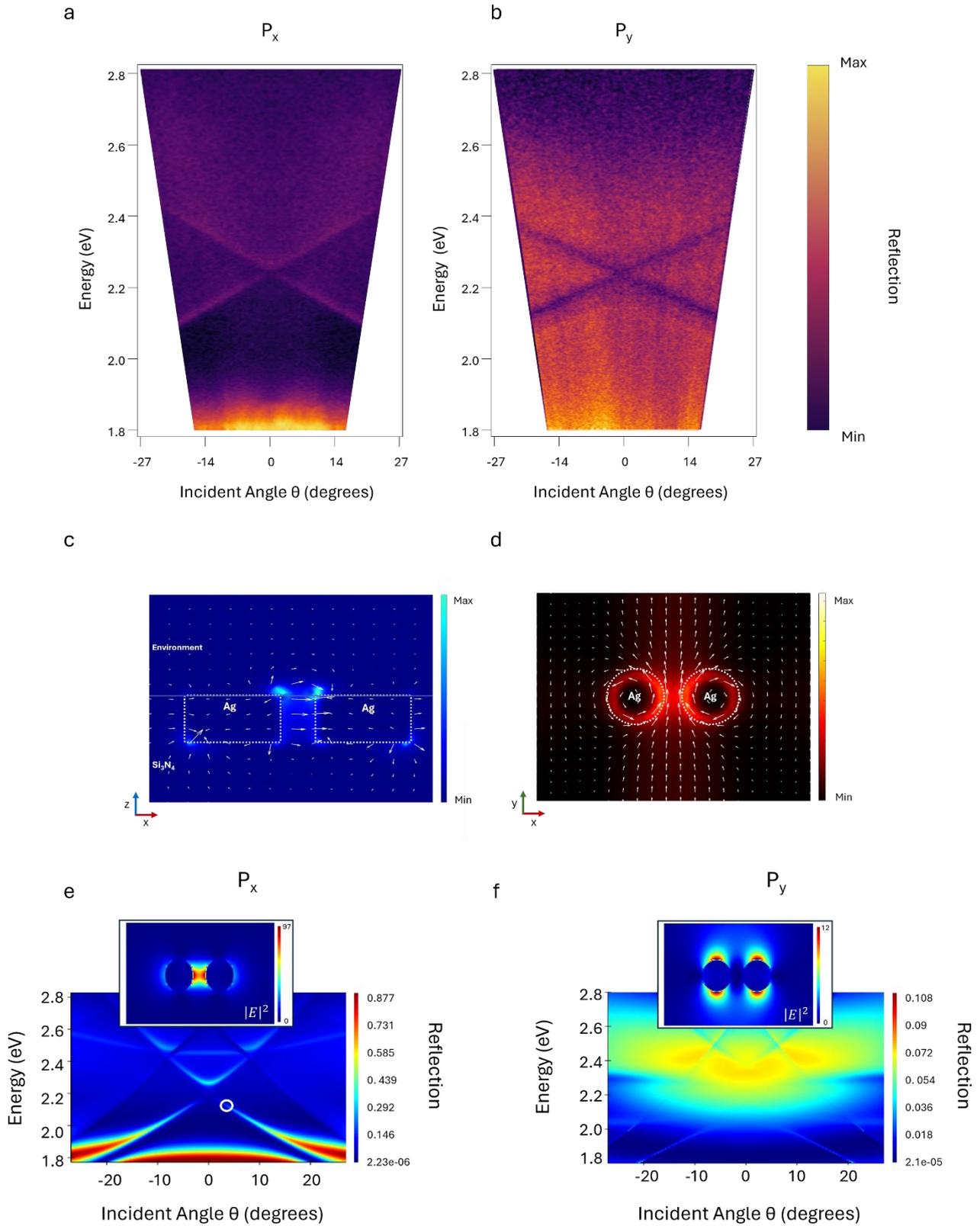

*Figure 2. a and b) Measured angular dispersion reflection maps as a function of the incident angle (ϑ°) for incident light polarization parallel ($P_x$) and normal ($P_y$) to the dimer axis. c and d) ) Electric and magnetic field vectors (arrows) and normalized intensities calculated at the quasi-BIC point for $P_x$. e) and f) Numerical simulation of the reflection far field maps. In the insets, the near-field distributions of the electric field are shown, corresponding to the quasi-BIC point circled in the map The value of $|E|^2$ shown is normalized to the incident electrical field $E_0 = 1\frac{V}{cm^2}$.*

To clarify the optical behavior of our system in the hybrid BIC generation along the dimers axis, we described the obtained dispersion and anti-crossing modes by a three-coupled harmonic oscillators model according to the following Hamiltonian:

$$H = \begin{pmatrix} E_{ph}^{+} & 0 & \frac{\Omega}{2} \\ 0 & E_{ph}^{-} & \frac{\Omega}{2} \\ \frac{\Omega}{2} & \frac{\Omega}{2} & E_{LSPR} \end{pmatrix}$$

where $E_{LSPR}$ is the plasmon energy, $\Omega$ is the Rabi splitting and $E_{ph}^{\pm}$ are the energy dispersions of the BIC-like photonic branches.

H gives rise to three hybrid modes: one upper branch (indicated as UB), one middle branch (MB) and one lower branch (LB). Notably, by fitting the modes position with the Hamiltonian eigenvalues, we found $E_{LSPR} \approx$ 1.9 eV and we extracted a Rabi splitting of $\Omega \sim$ 200 meV. In Fig. 3a are shown the polariton dispersions where the hybrid modes appear as a function of a color coded plasmonic/photonic fractions, calculated by the Hopfield coefficients for the branches around the plasmon energy. These fractions are overlapped to the corresponding measured reflection map (also shown in Fig.S5g without overlapping).
Interestingly, while the UB is affected to a lesser extent by the LSP position (exhibiting a more photonic character), a stronger coupling involves the MB and LB mode. Anyhow, the plasmonic/photonic content determines the coupling regime.

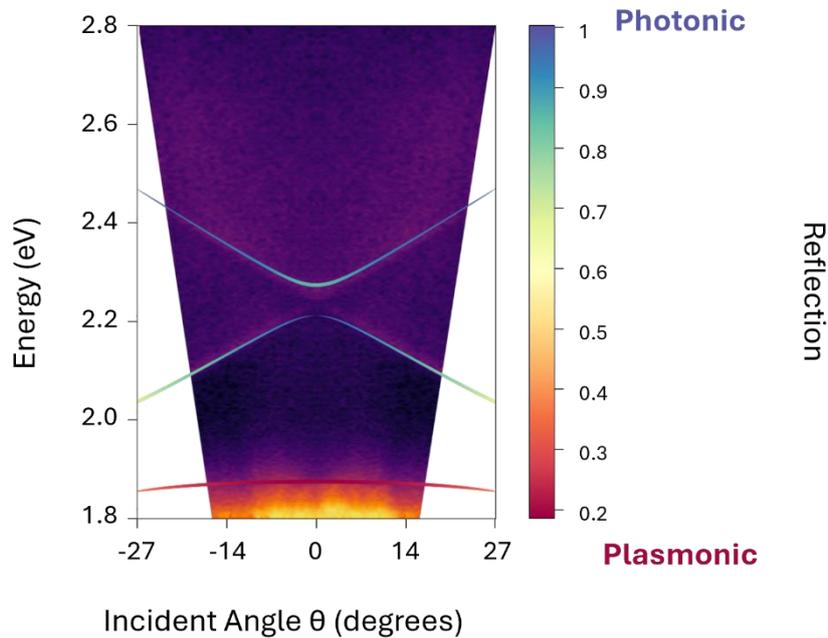

*Figure 3.) Measured reflection map along $P_x$ with overlapped color coded lines representing the eigenvalues calculated by diagonalizing the 3x3 Hamiltonian..*

Mode volume (field enhancement) versus Q factor

For a more in-depth analysis we extracted the theoretical Q-factors and the modal volumes of the hybrid quasi-BIC as a function of the DE value, varying the lattice period (LP) of the dimers along $P_x$ and producing a spectral scan of the photonic branches on the LSP mode, as shown in Fig. 4a (the corresponding simulated reflection maps are shown in Fig. S4). A definition of modal volume $V_{mod}$ can be given in terms of the electric energy density $u_E = \frac{1}{2}\varepsilon_r\varepsilon_0|\vec{E}|^2$ in which $\varepsilon_r$ is the relative permittivity, as

$$V_{mod} = \frac{\int_\Xi u_E \, dV}{u_E^{max}}$$

where $\Xi$ formally extends to infinity in the direction orthogonal to the array plane and $u_E^{max}$ is the maximum energy density within the same region of interest.[27] Considering the difficulty in calculating the mode volume we defined an effective modal volume $V_{eff}$ ranging between the bottom and the top surfaces of the NDs, and whose base area covers the entire unit cell of the array.[28,29]

The reflection spectra were fitted with a Lorentzian shape, and the full width at half-maximum close to the peak centre at the given λ was used to determine the Q-factor, which was calculated as $\frac{\lambda}{\Delta\lambda}$.

According to what is shown in Fig. 4a, smaller LP values induce a strong blueshift of the BIC mode, significantly increasing the DE with the LSP mode (weak coupling regime) and exhibiting a more photonic behavior, as a BIC-like diffractive order. As a result, the Q-factor values in this region are very high but the large modal volumes do not allow to reach an optimal Q/V ratio. On the other hand, large LPs enable the full spectral and spatial overlap between the two bare modes (strong coupling regime). Here, the prevailing plasmonic fraction leads to very small modal volumes but that's not sufficient to compensate the Q values decrease, which overall results in Q/V dropping.

Therefore, following these results, we defined an intermediate DE region for an optimum balance between large Q-factor and ultrasmall V coincident with the $P_x$ case in our experiment[15]. These results could be explained also by Hopfield coefficient values as calculated in Fig. S5. Indeed, a sufficiently low plasmonic fraction (16%), corresponding to a 0.5 DE value, ensures the hybrid BIC onset, with maximized Q/V ratio.

When switching the polarization along the y-axis, the optical behavior resembles the condition of strong coupling identified in Fig. 4a but with the distinction that in this case, as earlier mentioned and shown in Fig. 2b and d, the hybrid BIC generation is compromised due to reduced electric field strength of the single ND and the high LSP losses acting as an additional destructive leaky channel.

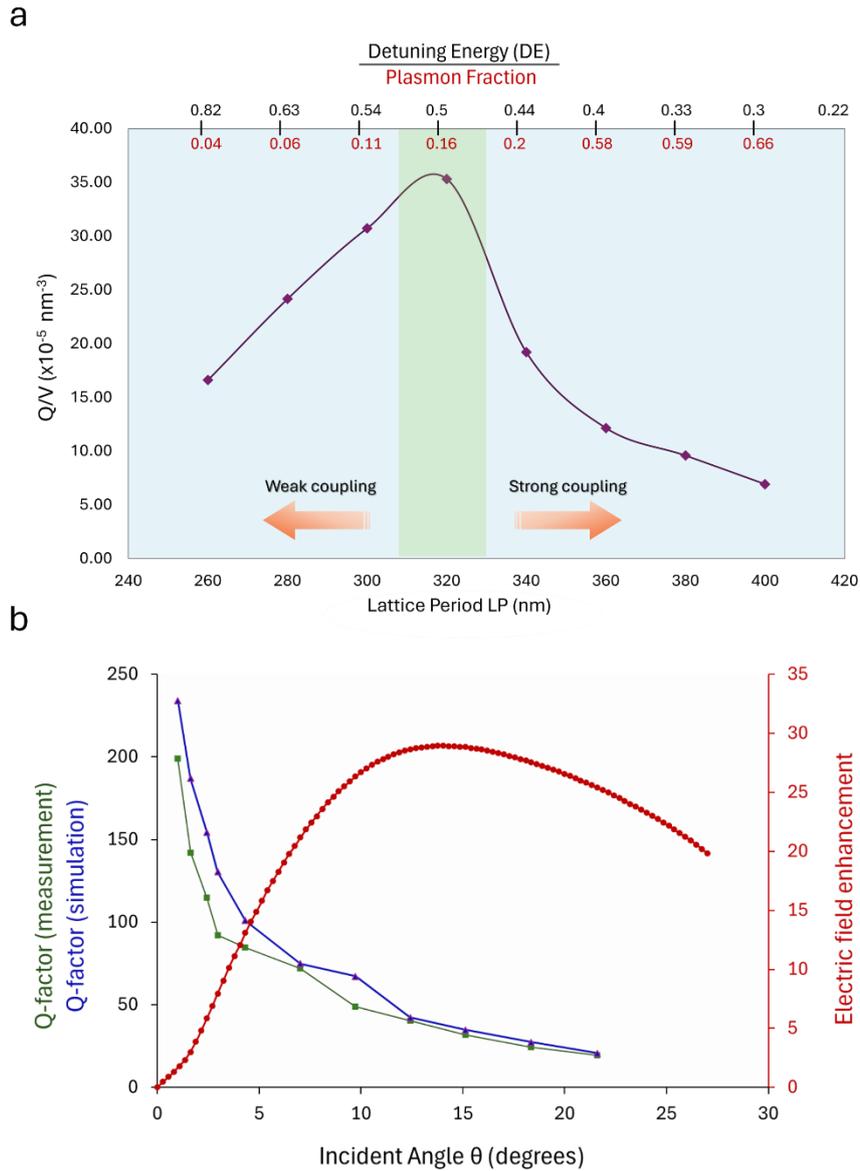

*Figure 4. a) Q/V ratio trend calculated for several lattice periods (along dimers axis) as a function of both the detuning energy (DE) between the LSP resonance and photonic BIC modes and the plasmon fraction derived from the Hopfield coefficients computation. The maximum value of Q/V is found at LP 320 nm (corresponding to a 0,5 eV DE and a 16% of plasmon fraction participating in the hybrid branch). b) Correlation among the simulated and measured Q-factor values and calculated electric field enhancement in the optimal DE region (320 nm lattice period along x-axis) as a function of the incident angle.*

To better evidence the photonic features of the hybrid modes, in Fig. 4b we show the experimental and simulated Q values of the hybrid MB as a function of the incident angle. Approaching $\vartheta=0$, the Q-factor exhibit a burst. However, the ability to confine and enhance light at the quasi-BIC resonance is very crucial for light-matter interaction. Here, the light confined at this regime, unlike regular guided modes below the light-line, can be excited by the free propagating plane wave. Therefore, a more quantitative insight into the plasmonic behavior of the coupled system (along $P_x$) can be gained by evaluating the trend of $\left|\frac{E}{E_0}\right|^2$ as a function of the incident angle which is shown in Fig. 4b. We noticed that the field enhancement hits a minimum at the BIC point and then increases up to 30 as we move away from $\vartheta=0$, while the Q-factor shows the opposite behavior reaching maximum value up to 200 at the quasi-BIC.

Such a study lays the foundation for the implementation of an innovative architecture as an advanced biosensing platform. Unlike other Γ-BIC based nanosystems, here a multifunctional approach opens the way towards the simultaneous enforcement of multiple independent targets, i.e. maximizing the light-matter coupling and reaching high performances in biosensing, by means of a specific detuning between the BIC energy valley and the LSPR that creates the conditions for losses suppression.

Hybrid BIC resonance tracking for biosensing

To test the potential of the proposed platform for biosensing applications, we have employed an engineered functionalized polymer with high yield coverage and specific biorecognition of target molecules of a neurodegenerative related biomarker. A schematic representation of the functionalization process is described in Fig. 5a while details can be found in experimental section. At first, the fabricated array (1) has been uniformly coated by an ultrathin (~10 nm) layer of polydopamine by self-polymerization. The presence of the ultrathin polymer does not alter how the hybrid BIC field interacts with molecules (2), as it does not change the plasmonic field distribution and the resulting refractive index shift does not induce a relevant detuning of the bare modes. Then we exploited the high density chemical moieties of the polymer for antibody covalent binding (3). At the end, as a proof-of-principle, we have tested the sensor's response to different concentration of TDP-43, a biomarker related to amyotrophic lateral sclerosis (ALS) and other neuropathies. To validate the repeatability of the experiment, four measurements were acquired on four different arrays. As shown in Fig. 5b, we observed a hybrid BIC resonance shift proportional to the analyte concentration ranging from 100 femtomolar to 100 zeptomolar and our sensor proved to be able to detect TDP-43 down to 100 zM. In Fig. S8 is shown more in detail the resonance shifts' identification procedure. To the best of our knowledge, this result goes beyond the interval accessible for this analyte through typical immunoassays [30–32], thus representing the first biosensing signature reaching this concentration level.

The sensing performances arise from the desired combination between the fundamental structural configuration and the ultrathin polymer layer that positively acts in a threefold manner: it prevents silver's chemical tarnishing [33], it guarantees the direct access to the high quasi-BIC field for the strong interaction with the target molecules and allows for a high-density immobilization of antibodies.

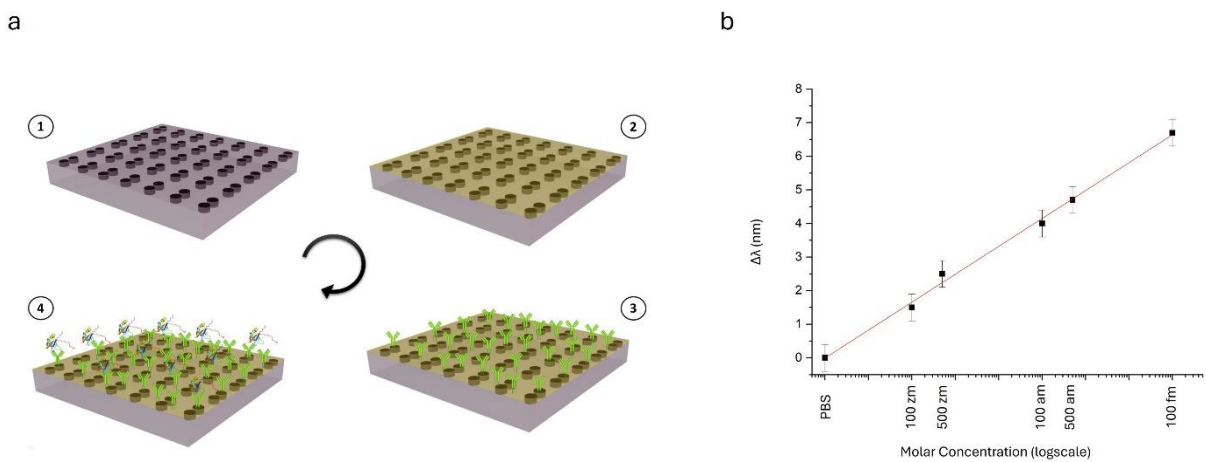

*Figure 5 a) Schematic representation of the functionalization process and biomolecule binding: 1- dimer based nanoarchitecture, 2- polymer layer coating, 3- antibody covalent binding, 4- target analyte TDP-43 covalent binding to its specific antibody. b) Relationship between the spectral difference between PBS (reference) and TDP-43 as a function of the molar concentration (LoD) detected by hybrid BIC resonance tracking.*

CONCLUSIONS

We envisaged a new functional optical platform combining photonic mode and LSP in a non-trivial 2D nanoarchitecture to build large light-matter interactions in a hybrid SP-BIC. The engineered building block sets the optical conditions to enter the coupling regime between LSP and quasi-guided modes, unlocking hybrid BIC states with high Q-factor values and small modal volume (boosted EM field enhancement). We reported that the hybrid BIC onset strongly depends on the spectral and spatial overlap of the bare modes but specially on the huge intensity of the plasmonic field. The latter is made possible only by the unique feature of the dimers nanogap (Px) with respect to the single ND case (Py). In addition, we identified a region of DE between the bare modes that grants the best trade-off in terms of Q/V by fine tuning of the plasmonic/photonic fractions involved in the hybridization, scanning the optical response of the nanosystem as a function of the LP. The proposed metasurface provides a further pivotal feature: the dimers' top surface is exposed to the environment, thus ensuring a direct access to the active surfaces which is a crucial requirement for high light-matter interaction. We have exploited this novel design in a proof-of-concept optical biosensing experiment and it proved to be highly effective in the detection of TDP-43, a biomarker related to ALS and other neuropathies. Strikingly, we were able to reach a 100 zeptomolar concentration, that constitutes an unprecedented level of bio-molecular sensing capability. All the above-mentioned characteristics can be well-tuned by tailoring the geometry of the nanostructures. Overall the configurational features of our nanoarchitecture provide multifunctional properties thus paving the way to versatile hybrid light matter optical devices for various applications such as lasing and biosensing

EXPERIMENTAL SECTION

*Matherials and methods:* Dopamine hydrochloride ($(HO)_2C_6H_3CH_2CH_2NH_2HCl$, powder, Sigma-Aldrich, USA), and polyclonal antibody anti-TDP-43 (liquid, Proteintech, USA) were used as received. Stock solution of TDP-43 (5 mM) was prepared in phosphate buffer solution at pH 7.4 and stored at -20° C if not used. Standard stock solution of dopamine hydrochloride (0.1 mg/mL) was prepared in Tris-HCl buffer solution (10 mM, pH 8.5). Phosphate-buffered saline (PBS), tris(hydroxymethyl) aminomethane hydrochloride (Tris-HCl) were prepared by using reagent-grade salts purchased from Sigma. All solvents, purchased from Sigma-Aldrich, are of the highest purity available. All aqueous solutions were prepared by using water obtained from a milli-Q Gradient A-10 system (Millipore, 18.2 MΩ cm, organic carbon content ≤ 4µg $L^{-1}$).

*Biomolecule binding*: Specific molecular detection using the fabricated metasurface has been realized, which involved three steps: surface activation, probe immobilization, and target capture. First of all, an ultrathin PDA layer was formed on the metasurface by keeping a solution of dopamine hydrochloride 0.1 mg/mL in Tris-HCl buffer at pH 8.5 overnight. In these conditions, the PDA catechol groups, responsible for adhesion on substrates, are oxidized to o-quinone, which are responsible for cross-linking [34]. The PDA coated metasurface was then rinsed with milli-Q water and dried under nitrogen flow. The bioconjugation reaction with TDP-43 was performed by exploiting the ability of PDA to immobilize biomolecules through reaction with $-NH_2$ groups, by simple incubation with an antibody solution 5 mM in buffer Tris-HCl at pH 8.5 for 1h. At alkaline pH, catechol and quinone functional groups present in the polydopamine coating are capable of covalent coupling to nucleophiles, thus allowing subsequent biomolecule immobilization through reaction between amines and polydopamine-coated substrates [35]. Then, the platform was rinsed with water and dried under nitrogen flow. Finally, the sensing performances of the fabricated nanosensor were investigated by incubating them with solutions of the un-tagged human recombinant TDP-43 protein at different concentrations ranging from 100 fM to 100 zM in PBS for 40 minutes. The device were then rinsed in milliQ water and finally dried. This measurement procedure was repeated on four different arrays and performed four times on each of them.

*Sample fabrication*. Two dimensional 80x80 silver nanodisks dimers with 70 nm diameter, 45 nm height and 30 nm gap were realized on a dielectric $Si_3N_4$ slab by electron beam lithography, with lattice periods 320x300 nm. The 210 nm thick layer of $Si_3N_4$ was evaporated via PECVD process onto a glass substrate. The sample was first cleaned in acetone and isopropyl alcohol . Then a 200 nm poly(methyl methacrylate) (PMMA) layer was spin coated at 3000 rpm and soft-baked at 180°C for 3 min. A 5 nm thick chrome layer was thermally evaporated onto the PMMA to prevent charge effects in the electron beam writing procedure. The arrays were written by a Raith 150 system at 26 pA beam current and 30 keV. After electron exposure, the Cr layer was completely removed by a ceric ammonium nitrate based wet etching for 40 s and rinsed in water. The exposed resist was development in MIBK:IPA solution in a 1:3 ratio for 3 min and rinsed in 2-propanol for 1 min. The sample was then treated with Inductively Coupled Plasma Etching (ICP) in order to recreate the design that was drawn on the resist onto the $Si_3N_4$ slab. After thermal evaporation of 40 nm of silver, a lift-off process was performed in an mr-Rem 500 remover solution (Microresist Technology) and rinsed in 2-propanol.

*Numerical Simulations*: Numerical simulations were performed by exploiting an FTDT and RCWA based software (Ansys Lumerical). Considering the array system investigated, all parametric studies have been performed by simulating an elemental unit cell with an Ag disk dimer embedded in a $Si_3N_4$ waveguide with the top surface exposed to the environment and employing periodic boundary conditions in the array to get energy–momentum reflection maps $E(k_x, k_y = 0)$ for different simulation designs. The glass substrate was not included in the simulations. For what concerns the optical properties of the plasmonic metal involved, we considered the Johnson and Christy formula for the permittivity dispersion of silver. From the environmental point of view, the disks have been considered exposed to a non-dispersive dielectric medium with refractive index n = 1.4 to simulate the effect of the 10 nm layer of polymer.

*Optical characterization*. The sample was characterized by reflection measurements on an home-made confocal setup comprising an optical microscope (Zeiss AxioScope A1) coupled to a spectrometer. Light from a tungsten lamp was focused on the sample by an adjustable numerical aperture condenser (NA from less than 0.1 to 0.95). Reflected light was collected with a 10x 0.95 NA objective lens. Subsequently, the light passes through a three lenses system: the first reconstructs the real space, the second collimates the light beam, and the third lens refocuses the image in the real space. The light was then directed to a Hamamatsu Orca R2 CCD camera and a 150 mm spectrometer. By using all the three lenses combined with an additional pinhole, the image can be selected in space. Removing the intermediate lenses, instead, Fourier space imaging had been obtained. An adjustable pinhole was used to select the array area in the focal plane.

Acknowledgements

We acknowledge financial support under the National Recovery and Resilience Plan (NRRP), Mission 4, Component 2, Investment 1.1, Call for tender No. 1409 published on 14.9.2022 by the Italian Ministry of University and Research (MUR), funded by the European Union – NextGenerationEU– Project Title  Chiral Bound States IN the Continuum by Shallow 3D Plasmonic SPIRal MEtacrystal (INSPIRE) – CUP

B53D23024270001 - Grant Assignment Decree No. 1380 adopted on 01/09/2023 by the Italian Ministry of Ministry of University and Research (MUR).

The work was supported by the Italian Ministry of Research (MUR) in the framework of the National Recovery and Resilience Plan (NRRP), "NFFA-DI" Grant (CUP B53C22004310006), "I-PHOQS" Grant (CUP B53C22001750006) and under the complementary actions to the NRRP, "Fit4MedRob" Grant (PNC0000007, CUP B53C22006960001), "ANTHEM" Grant (PNC0000003, CUP B53C22006710001), funded by NextGenerationEU. U. The work was also supported by "Tecnopolo per la medicina di precisione" (TecnoMed Puglia) – Regione Puglia: DGR no. 2117 del 21/11/2018 CUP: B84I18000540002 and by the National project "Developing National And Regional Infrastructural Nodes Of Dariah In Italy – DARIAH" (CIR01_00022)

# Hybrid plasmonic Bound State in the Continuum entering the zeptomolar biodetection range

Elena Clabassi[∥,1,2], Gianluca Balestra[∥,1,2], , Giulia Siciliano[1], Laura Polimeno[1], Iolena Tarantini[2], Elisabetta Primiceri[1], David Maria Tobaldi[1], Massimo Cuscunà[1], Fabio Quaranta[3], Adriana Passaseo[1], Alberto Rainer[1,5], Silvia Romano[4], Gianluigi Zito[4], Vittorianna Tasco[1‡], Marco Esposito[1,2]*

1. CNR NANOTEC Institute of Nanotechnology, Via Monteroni, Lecce 73100, Italy

2. Department of Mathematics and Physics Ennio De Giorgi, University of Salento, Lecce 73100, Italy

3. CNR IMM Institute for Microelectronics and Microsystems, Via Monteroni, Lecce 73100, Italy

4. CNR ISASI Institute of Applied Sciences and Intelligent Systems, Naples, Italy

5. Department of Engineering, University Campus Bio-Medico di Roma, via Álvaro del Portillo 21, 00128 Rome, Italy

*E-mail: marco.esposito@nanotec.cnr.it

∥ E.C. and G.B. contributed equally to this work.

‡Vittorianna Tasco is currently seconded at the European Research Council Executive Agency of the European Commission. Her views expressed in this paper are purely those of the writer, may not in any circumstance be regarded as stating an official position of the European Commission.

S1. Photonic BIC in a dielectric slab

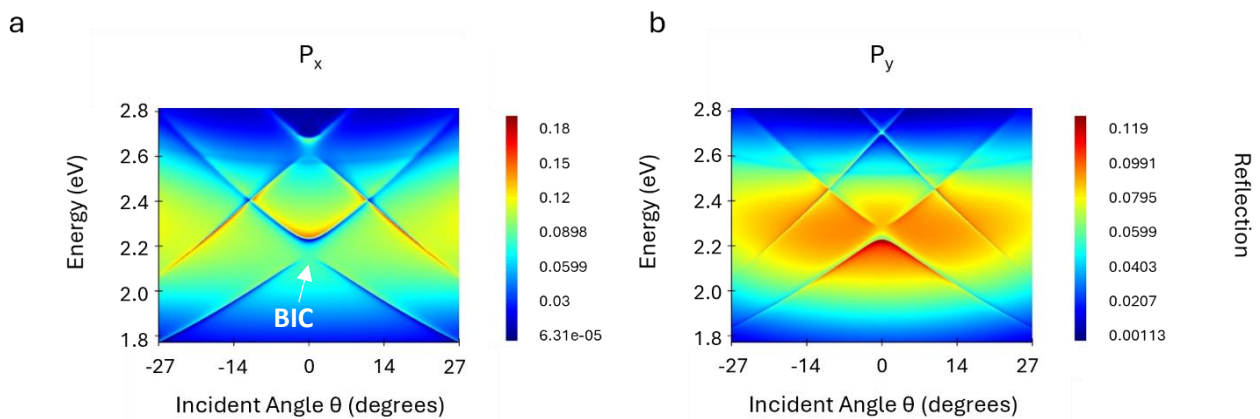

In **Figure S1** we display the the comparison between $P_x$ and $P_y$ angular dispersion describing the emergence of a purely photonic BIC state from a dielectric $Si_3N_4$ waveguide with a 2D subwavelength grating based on dimers nanoholes. In this scenario, we are able to identify the guided modes because they get diffracted into the radiative continuum thanks to their coupling with Bragg scattering. Along $P_x$ (Fig.S1a) the engineered tuning of the nanoholes grating leads to Bloch resonances exhibiting a gap opening and the formation of symmetry-protected BIC at the Γ point of the momentum space. The BIC branch has a maximum in intensity values and appear as dark at normal incidence whereas the "lossy" branch reaches a minimum and is bright. Conversely, along $P_y$ (Fig.S1b) we observe a symmetry reversal and the emergence of the BIC mode in the lower branch of the anti-crossed GMs. We note how the gap formed between the hybrid modes is wider in the $P_x$ configuration in which also the intensity of the electrical field is higher due to the effect of confinement enhanced by the gap of the dimer.

## S2. Plasmonic grating without waveguide

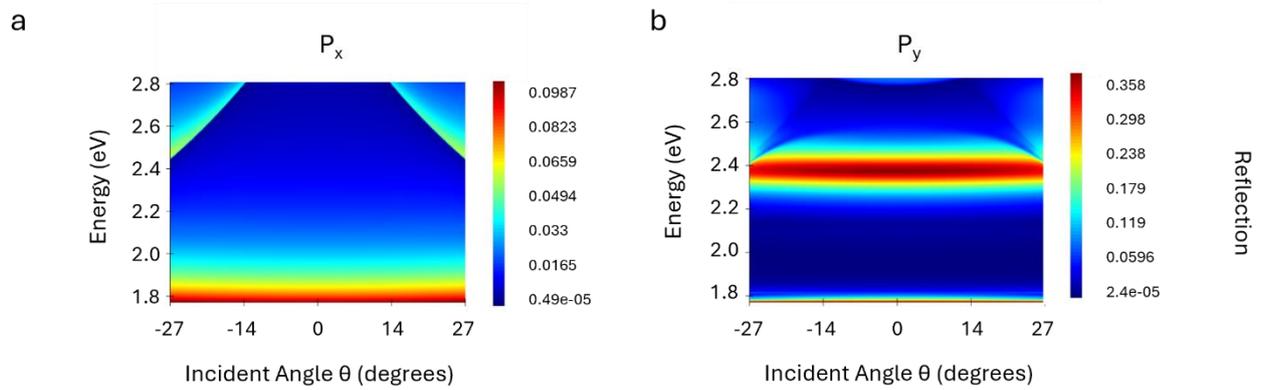

In **Figure S2** are shown the simulations of the angle-resolved reflectance for Ag NDs calculated in a 1.4 refractive index environment (polymer solution) for both the polarizations discussed in the main text. The spectra display the characteristic dispersion branches of the couple of diffractive orders arising from light refraction with the geometry. Additionally, the reflection maps show a single broad LSP resonance: for the dimers configuration it is found at E= 1.7 eV (Fig.S2a), whereas the one related to the single ND appears at E = 2.3 eV (Fig.S2b).

## S3. LP Scanning

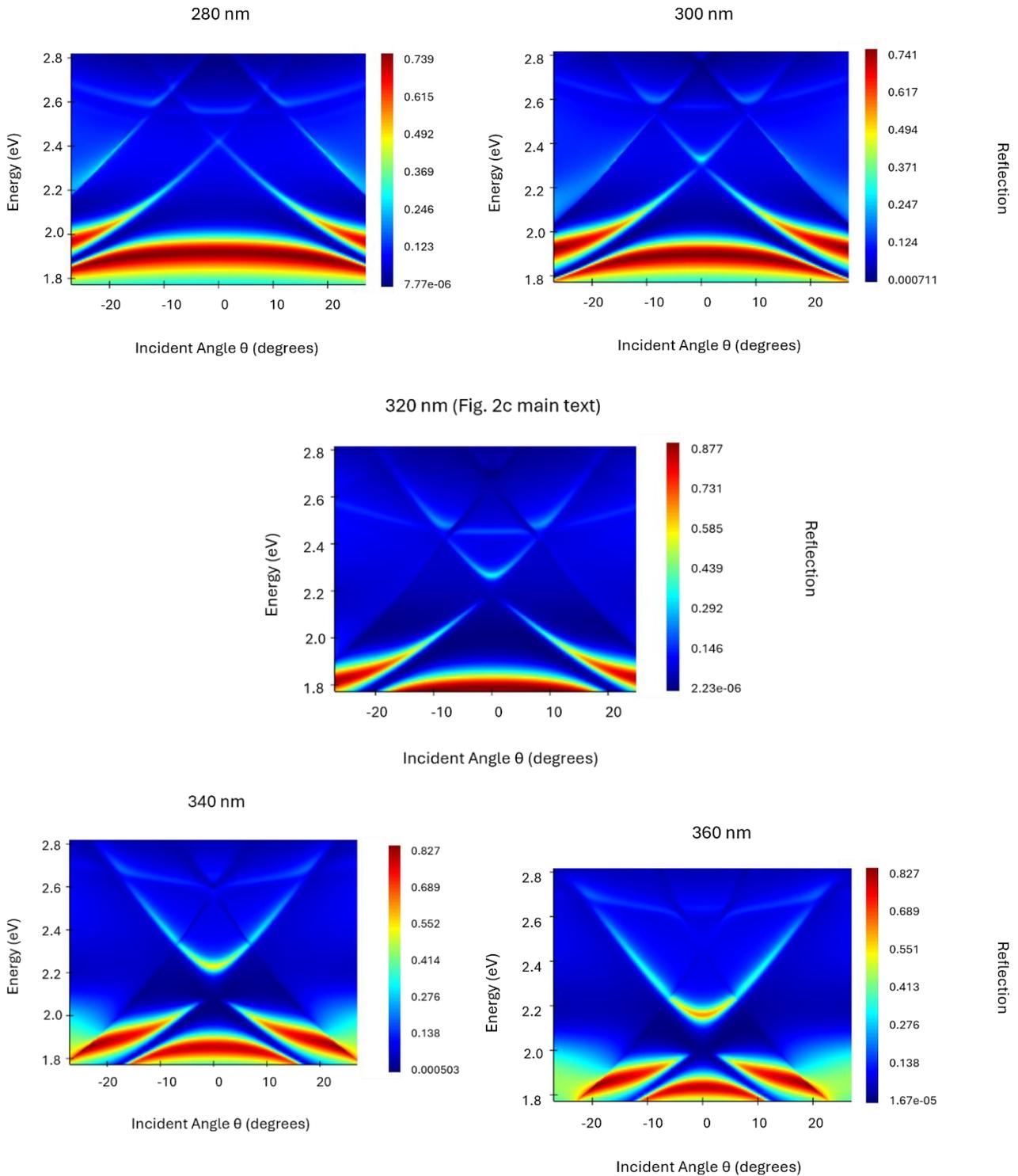

In **Figure S3** we display the simulated reflection maps related to the LP scanning along the dimer axis producing a spectral scan of the BIC-like photonic branches on the LSP mode.

## S4. Plasmonic grating on top of the waveguide

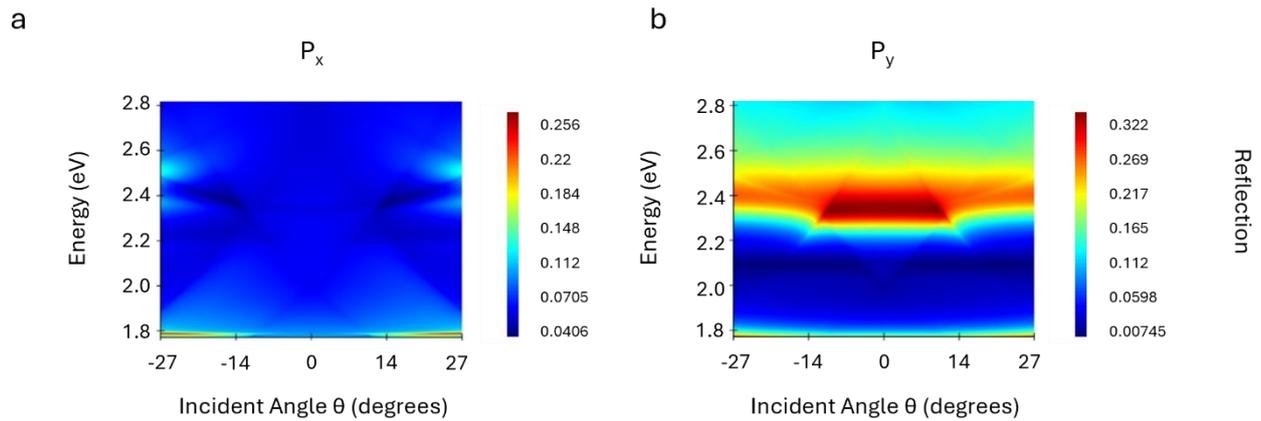

**Figures S4** displays the angle-resolved reflection maps for a $Si_3N_4$ waveguide with a 2D array of silver dimers placed on top and simulated in a 1.4 refractive index environment. Along the dimers polarization axis one can see the dispersions related to diffractive modes of the first and second order together with the LSP resonance related to the dimer at 1.7 eV. Switching to $P_y$, one can still find the diffractive modes as mentioned for $P_x$, but the LSP resonance is blue-shifted since corresponding to the single ND component. In neither case the hybrid BIC is generated.

## S5. Perturbation method

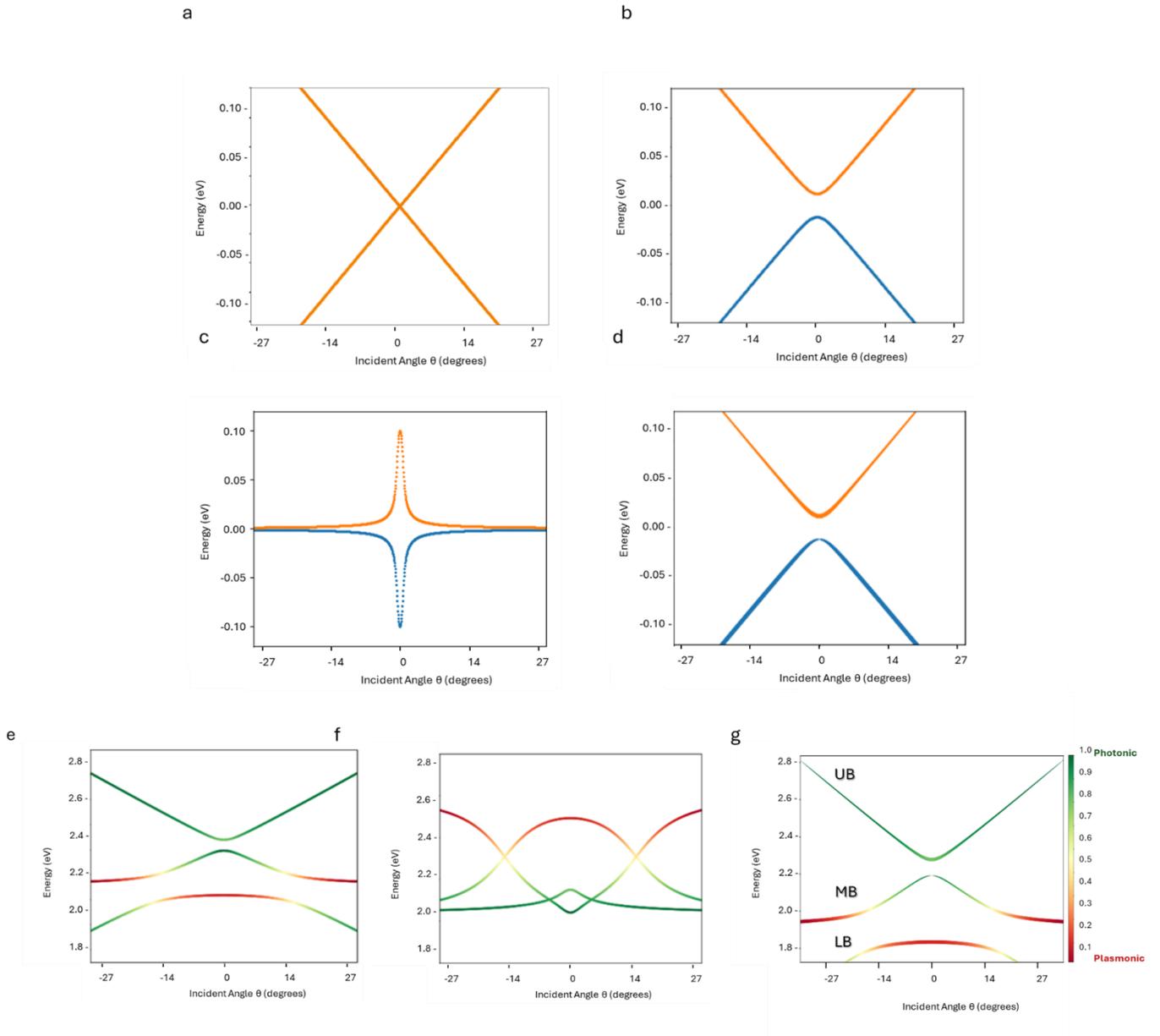

In the main text we have described an analytical model for the coupling regime between the photonic BIC and the LSP resonance based on the calculation of the Hamiltonian for three-coupled oscillators. In **Figure S5** we aim to illustrate the steps that generate the BIC mode from the guided modes, eventually leading to Fig.3a (main text) where the hybridization with the plasmonic component is shown. First, we investigated the behaviour of the fundamental guided modes in the slab. The results are shown in Fig.S5a: presenting only band folding effects, the linear dispersions of the two uncoupled GMs cross one another. Then, we introduced the modifications due to a periodic grating: this induces a diffractive coupling between the modes and consequently the opening of a gap between the resulting Bloch resonances and the formation of SP-BIC at the Γ point of the momentum space. We analyzed this scenario through a two-coupled oscillators system as reported in Eq. 1 below. Here, the term $\bar{\bar{\gamma}}$ indicates a matrix whose components take into account the radiative losses in the coupling. In Fig.S5b-c are shown the real and imaginary part, respectively, of the related dispersions. Also, in Fig.S5d is reported the overall dispersions plot to better highlight the formation of the BIC state.

$$H = \begin{pmatrix} E_{ph}^+ & g \\ g & E_{ph}^- \end{pmatrix} + i\bar{\bar{\gamma}} \qquad \text{Eq. 1}$$

Now, taking into consideration [1], the second step of perturbation consists in the coupling between Bloch resonances and the LSP resonance to form an hybrid plasmonic-photonic SP-BIC. As discussed in the main text (Fig.3a), we approximated this system with a three-coupled oscillators model in which the LSP resonance couples to both the photonic modes relating to the dielectric grating. In Fig. S5e-g are shown the real, imaginary and total components of the dispersions characteristic of the three polariton branches ($\Omega = 260$ meV, $\gamma = 0.14\ meV$).

## S6. Q-bic shift tracking

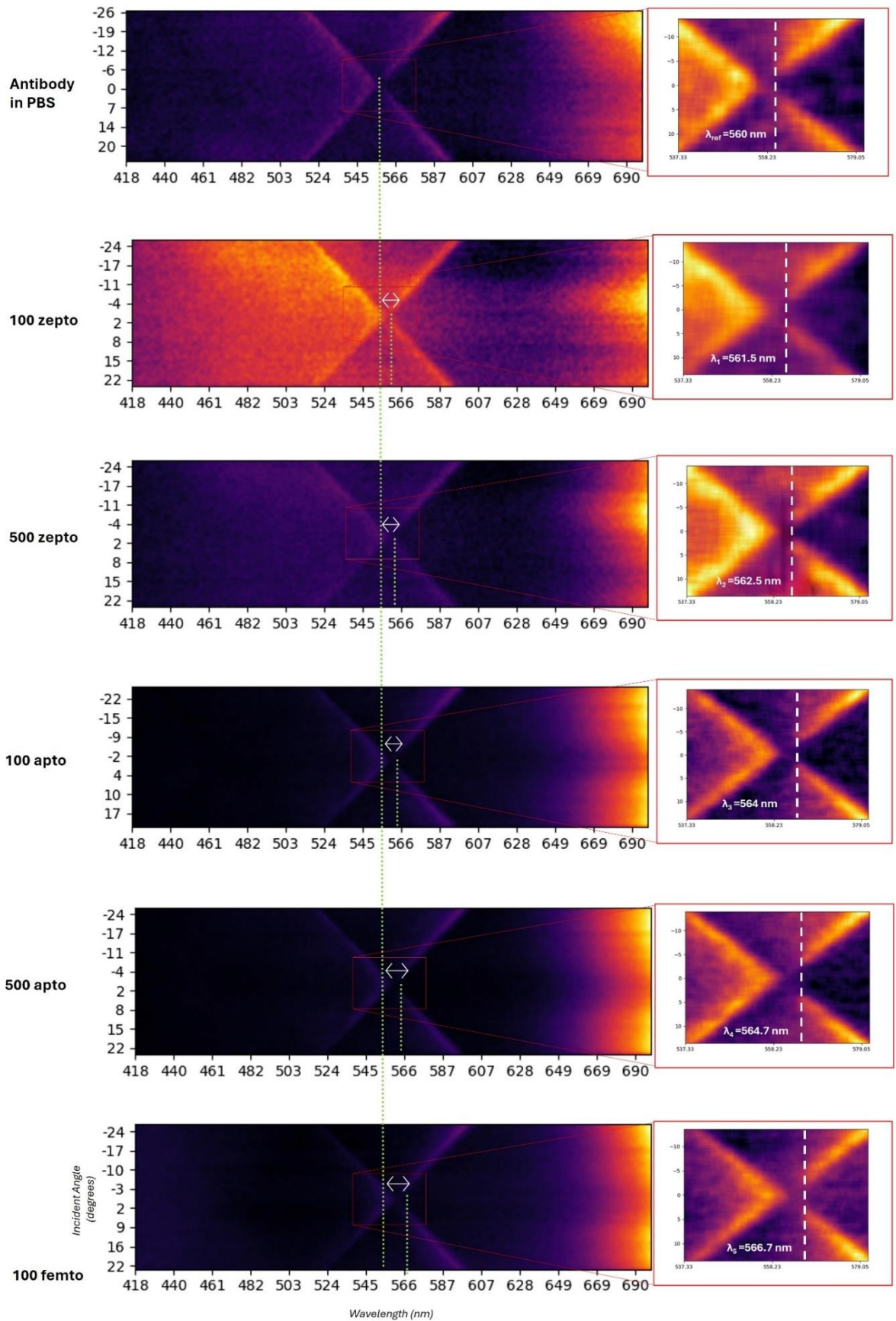

The measured dispersion maps for the antibody and the range of TDP-43 concentrations are shown in **Figure S6**. We considered the quasi-BIC states' resonance in order to determine the resonance shift attributed to the analyte detection. It's possible to widely visualize on the maps that these states have shifted in relation to the antibody's one, which is used as a reference. The quasi-BICs can be examined in greater detail from the inlets, and each measurement's associated wavelength is provided. As expected, the highest concentration exhibits the largest resonance shift, but remarkably, a 1.5 nm shift was also present at the 100 zeptomolar concentration, suggesting an unparalleled detection limit for these kinds of biosensing nanostructures.

S7. Transactive response (TAR) DNA-binding protein 43 (TDP-43)

A variety of disorders known as neurodegenerative diseases (NDDs) are distinguished by continuing loss and selective malfunctioning of neurons, glial cells, and neural networks in the brain and spinal cord. One of the most common is amyotrophic lateral sclerosis (ALS), a uniformly lethal progressive degenerative disorder of motor neurons that also overlaps with frontotemporal lobar degeneration (FTLD) clinically, morphologically, and genetically[2]. Currently ALS is clinically diagnosed after a typical diagnostic delay of almost a year from the onset of symptoms, by which time the disease is well-established and the window for potential therapy may have closed. In order to address the growing incidence of this disease, it is imperative to create selective and precise biomarkers for early diagnosis and treatment development. In this regard, Transactive response (TAR) DNA-binding protein 43 (TDP-43) has been acknowledged as a potentially useful biomarker. In fact, TDP-43 builds up in nerve cells in most cases of Tau-negative frontotemporal lobar degeneration (FTLD) and almost all cases of ALS [3,4] Familial and sporadic ALS and FTLD cases have been linked to the development of TDP-43 mutations. Using mass spectrometry [5], Western Blot, and ELISA test [6,7], it was discovered that patients with ALS and FTLD had elevated levels of TDP-43 protein in their cerebrospinal fluid (CSF) and plasma. TDP-43 was recently found in CSF at a concentration of less than 0.49 ng/mL via an ELISA test [8] and a detection limit of 0.5 ng/mL [9] has been reached in serum utilizing an electrochemical sensor. However, most of these methods suffer from drawbacks such as high cost and variability [10], complex apparatus requirements, higher limit of detection (LOD), poor reproducibility, and dependence on operator. Furthermore, TDP-43 levels are often below the detection limit of standard immunoassays in a significant portion of both patients and healthy persons [6,7]. Thus, for a more precise validation of this biomarker and, as a result, for the early diagnosis of ALS and FTLD, more sensitive tests are needed to quantify lower TDP-43 concentrations present in complex fluids.